\pdfoutput=1
\documentclass{ceurart}

\usepackage{graphicx}
\usepackage{listings}
\begin{document}

\copyrightyear{2026}
\copyrightclause{Copyright for this paper by its authors.
  Use permitted under Creative Commons License Attribution 4.0
  International (CC BY 4.0).}

\conference{ISWC 2026 Posters and Demos Track, October 27--29, 2026, Bari, Italy}

\title{Adapting Knowledge Graphs for Behavior Denoising in Sequential Recommendation}

\author[1]{Zichun Jin}
\fnmark[1]

\author[1]{Zihan Zhou}
\fnmark[1]

\author[1]{Yinan Liu}
\cormark[1]

\author[1]{Bin Wang}

\author[1]{Xiaochun Yang}

\address[1]{Northeastern University, Shenyang, China}

\cortext[1]{Corresponding author.}
\fntext[1]{These authors contributed equally.}

\begin{abstract}
Sequential recommendation predicts the next item from a user's interaction history, but not every interaction is equally informative. Real logs combine persistent preferences with temporary needs, exploration, and incidental behavior, so some interactions can distort history representations or provide unreliable supervision. Existing denoising methods judge such interactions mainly from co-occurrence, order, or model predictions, without explicit evidence from relations between items. Knowledge graphs (KGs) offer this evidence, but item popularity, graph degree, uneven coverage, and widely shared entities can inflate connectivity and bias reliability estimates.
Here we present \textsc{AdaptedKG}, which derives calibrated KG evidence for each training example without adding graph representations to the recommendation model. It first compares the observed context with structurally matched alternatives to identify relational paths that are unusually prominent and uses them to build a local KG view. It then compares each interaction with structurally matched reference items to calibrate its support within that view. The resulting retention coefficients gate historical representations and reweight target losses. All sample-specific scores are computed offline using training
interactions and a fixed KG, so the backbone remains unchanged and no
KG access is required at inference. Experiments show gains with a standard sequential recommender and multiple behavior-denoising sequential recommenders.
\end{abstract}


\maketitle

\section{Introduction}
Sequential recommendation relies on a user's interaction history to
predict the next item~\cite{DBLP:conf/icdm/KangM18,
DBLP:conf/www/QuadranaJC19}, yet the interactions in this history do not
provide equally reliable evidence of user preference~\cite{DBLP:conf/ijcai/SunWSY21}. Real-world logs
reflect a mixture of long-term preferences, short-term needs,
exploration, and incidental behavior, making individual interactions
unevenly valuable for sequence modeling~\cite{DBLP:conf/recsys/ChenLPWYLZWY22}. An interaction weakly
associated with other interactions in the same sequence may both
disrupt historical representations and provide unstable supervision
when used as a training target~\cite{DBLP:conf/ijcai/SunWSY21}. Existing denoising methods mainly rely
on co-occurrence, order, or model predictions to identify such
interactions~\cite{DBLP:conf/www/LinWCR0YR0R23,  DBLP:conf/nips/SunWSYW23, DBLP:conf/icde/ZhangH00TS24}, but lack evidence from explicit inter-item relations for
making this judgment.

Knowledge graphs encode entities and relations connecting items and
    therefore offer a potential means of filling this gap~\cite{DBLP:journals/csur/HoganBCdMGKGNNN21, DBLP:journals/tkde/GuoZQZXXH22}. However, richer
path connectivity in the graph does not necessarily imply more relevant
relations between items~\cite{DBLP:conf/www/ShomerJ0T23}. Items with higher popularity, higher KG
degree, or more extensive coverage naturally admit more paths, while
entities shared by many items can also induce broad connections. Using
these connections directly may bias reliability estimation toward
richly connected items and common relational patterns. Whether a path provides distinctive support for the current interaction
must therefore be assessed against structurally comparable references.

Based on this idea, we propose \textsc{AdaptedKG}. Rather than injecting
KG representations into the recommendation backbone, it uses two
structural-matching stages with distinct roles that are executed
sequentially. These stages convert connections in the global KG into
KG-consistency evidence for individual behavioral examples and derive
conservative retention coefficients to modulate training signals. In
the first stage, alternative contexts with similar structural
attributes characterize the background distribution of relational
paths. The method retains and weights relational patterns that are more
prominent in the current context relative to the matched background,
forming an example-specific local KG view. With this view fixed, the
second stage uses reference items with similar structural attributes to
calibrate the support that these relational patterns provide for the
evaluated interaction. The resulting retention coefficients gate
historical interaction representations and reweight target losses,
respectively. All KG adaptation and sample-specific scoring are performed offline
using training interactions and a fixed KG, without changing the
backbone architecture or requiring KG access at inference. Experiments show that
\textsc{AdaptedKG} yields improvements when applied to a
standard sequential recommendation model and multiple
behavior-denoising models.

\section{Related Work}
Behavior denoising improves sequential recommendation by correcting
unreliable interactions or reducing their influence during training.
STEAM~\cite{DBLP:conf/www/LinWCR0YR0R23} learns item-wise keep, delete, and insert operations
from synthetically corrupted sequences. BirDRec~\cite{DBLP:conf/nips/SunWSYW23} uses
bidirectional recommender predictions to rectify unreliable histories
and targets with theoretical error guarantees. SSDRec~\cite{DBLP:conf/icde/ZhangH00TS24}
exploits interaction-derived inter-sequence relations to guide
self-augmentation and hierarchical denoising. Despite their different
mechanisms, these methods obtain correction evidence from interaction
logs, whether through synthetic supervision, recommender predictions,
or learned behavioral relations. These log-derived signals do not directly provide external typed
relational evidence for assessing an interaction's contextual support.
KG-based recommenders commonly integrate graph-derived representations
into prediction models~\cite{DBLP:journals/tkde/GuoZQZXXH22};
\textsc{AdaptedKG} instead uses structurally matched KG evidence to
derive example-conditioned retention coefficients offline, before
recommendation optimization.

\section{Methodology}
Raw KG connectivity is not directly comparable across examples. A context containing popular or high-degree items may contain many common paths, while an easily connected candidate may receive high support for the same reason. \textsc{AdaptedKG} handles these effects in sequence: matched-null contexts select paths that stand out from comparable contexts; with those paths fixed, matched references convert support into an interaction-level retention coefficient.

\subsection{Example-Conditioned Interaction Reliability}

For a training example $\mathcal D_s=(H_s,y_s)$, let $H_s=(x_{s,1},\ldots,x_{s,L_s})$ contain only interactions preceding $y_s$. Let $R(v;C)\in[0,1]$ be the retention coefficient assigned to $v$ from context $C$:
\begin{equation}
r^H_{s,j}=R(x_{s,j};H_s^{-j}),
\qquad
r^T_s=R(y_s;H_s),
\label{eq:role-reliability}
\end{equation}
where $H_s^{-j}$ excludes position $j$. When scoring $x_{s,j}$, this avoids direct self-inclusion; the target $y_s$ is not in $H_s$ and is scored against the complete prefix. The coefficient $r^H_{s,j}$ scales the historical embedding, while $r^T_s$ scales the target loss.

\subsection{Local KG Adaptation with a Matched Null}

Let $C=(c_1,\ldots,c_n)$ be an ordered context. A directed typed two-hop pattern $p=(r_1,c,r_2)$ links an ordered item pair via connector $c$; $\phi_p(u,v)$ indicates whether $(u,v)$ instantiates $p$. High coverage in $C$ does not by itself make $p$ informative, because similar coverage may also occur in contexts with comparable structural attributes. We therefore form matched-null contexts $\{C^{0,b}\}_{b=1}^{B_A}$ by replacing each position with an item of similar training popularity, KG degree, and linkage status while preserving length and order. For a value $t$ and nonempty multiset $\mathcal A$, $Q(t;\mathcal A)$ records the fraction of values below $t$, with ties contributing one half:
\begin{equation}
Q(t;\mathcal A)=\frac{1}{|\mathcal A|}
\sum_{a\in\mathcal A}
\left[\mathbb I(a<t)+\tfrac12\mathbb I(a=t)\right].
\label{eq:midrank}
\end{equation}
The observed coverage and its matched-null weight are
\begin{equation}
\begin{aligned}
\kappa_p(C)
&=\frac{1}{n(n-1)}
\sum_{\substack{1\leq a,b\leq n\\a\neq b}}
\phi_p(c_a,c_b),\\
\alpha_p(C)
&=\left[\kappa_p(C)-\operatorname{median}_{b}
\kappa_p(C^{0,b})\right]_+
Q\!\left(\kappa_p(C);
\{\kappa_p(C^{0,b})\}_{b=1}^{B_A}\right).
\end{aligned}
\label{eq:local-adaptation}
\end{equation}
The first line measures the fraction of ordered position pairs covered by $p$. In the second, the bracketed term keeps the path only when its coverage exceeds the matched-null median, while $Q$ gives more weight when that coverage ranks higher among the null values. The weighted paths with $\alpha_p(C)>0$ form $\mathcal P_C^*$ and define the local KG view $\mathcal G_C^*$.

\subsection{Matched Support Calibration and Training Adaptation}

Even after path filtering, an easily connected item may match many retained paths. We therefore match references $\{z_b\}_{b=1}^{B_R}$ to $v$ by training popularity, KG degree, and linkage status. They are scored only after $\mathcal G_C^*$ is fixed, so they cannot change the selected paths:
\begin{equation}
\begin{aligned}
S_K(v\mid C)
&=\frac{
\sum_{a=1}^{n}\sum_{p\in\mathcal P_C^*}
\alpha_p(C)\phi_p(c_a,v)
}{n\sum_{p\in\mathcal P_C^*}\alpha_p(C)},\\
R(v;C)
&=\min\!\left\{1,
2Q\!\left(S_K(v\mid C);
\{S_K(z_b\mid C)\}_{b=1}^{B_R}\right)\right\}.
\end{aligned}
\label{eq:support-calibration}
\end{equation}
$S_K(v\mid C)$ is the average fraction of retained path weight connecting $v$ to the context. The second line compares this score with the matched-reference scores:
a mid-rank percentile below one half gives $R(v;C)=2Q<1$, while one
at or above one half gives $R(v;C)=1$.

Each historical embedding is multiplied by its coefficient, $\widetilde{\mathbf e}_{s,j}=r^H_{s,j}\mathbf e(x_{s,j})$, before sequence encoding. For $\ell_s=\operatorname{CE}(f_\theta(\widetilde H_s),y_s)$ and batch $\mathcal B$, the target-weighted loss is $\sum_{s\in\mathcal B}r_s^T\ell_s/\sum_{s\in\mathcal B}r_s^T$ when the denominator is positive, and the unweighted mean otherwise.

If $n<2$, a valid matched background or reference set cannot be constructed, or $\mathcal P_C^*=\varnothing$, the KG channel uses $R(v;C)=1$ and skips undefined quantities. Global matching statistics use the training partition; sample-specific retention coefficients are computed offline using training interactions and a fixed KG, and are detached from recommendation optimization.

\section{Experimental Results}
\paragraph{Experimental setup.}
We evaluate KG calibration and sequential recommendation on Steam Games,
which contains 25,389 users, 4,089 items, 328,278 interactions, and
462,016 KG triples over six relations. We use SASRec~\cite{DBLP:conf/icdm/KangM18},
STEAM~\cite{DBLP:conf/www/LinWCR0YR0R23},
BirDRec~\cite{DBLP:conf/nips/SunWSYW23}, and
SSDRec~\cite{DBLP:conf/icde/ZhangH00TS24}, all of which are implemented within the RecBole framework~\cite{DBLP:conf/cikm/ZhaoMHLCPLLWTMF21}. For each backbone and its \textsc{AdaptedKG}-enhanced counterpart, we use identical data splits, candidate sets,
and backbone hyperparameters. Retention coefficients and interaction-derived matching statistics
are computed using training interactions only.

\begin{figure}[t]
  \centering
  \begin{minipage}[t]{.475\linewidth}
    \centering
    \includegraphics[width=\linewidth]
      {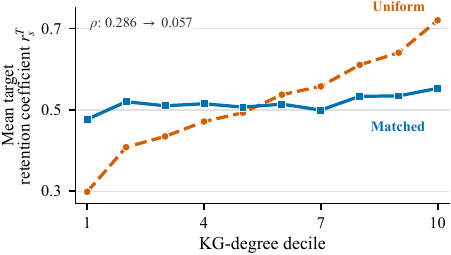}
    \smallskip
\hspace*{.14\linewidth}%
\makebox[.86\linewidth][c]{\small (a) KG degree}
  \end{minipage}%
  \hfill%
  \begin{minipage}[t]{.475\linewidth}
    \centering
    \includegraphics[width=\linewidth]
      {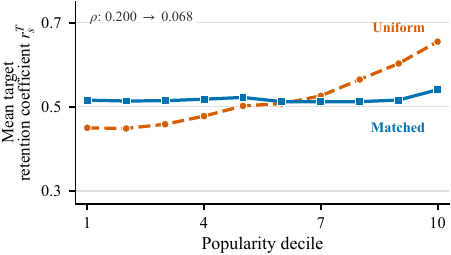}
    \smallskip
    \hspace*{.14\linewidth}%
    \makebox[.86\linewidth][c]{\small (b) Item popularity}
  \end{minipage}
  \caption{\textbf{Matched-reference sampling attenuates the structural
  dependence of interaction reliability on Steam.}
  Mean target retention coefficient $r_s^T$ across deciles of
  (a) KG degree and (b) item popularity under uniform and matched
  reference sampling.}
  \label{fig:kg-calibration}
\end{figure}

\paragraph{KG-side calibration.}
We isolate reference calibration by fixing the target queries, local path
views, and number of references while changing only the reference sampler.
The substantially weaker correlations under matched sampling
(Figure~\ref{fig:kg-calibration}) indicate that the estimated retention
coefficients are less coupled to item-side structural exposure.

\begin{table}[pos=h]
  \caption{Steam Games results ($\times10^{3}$).
  Each entry in panel (a) reports the backbone result followed by its \textsc{AdaptedKG}-enhanced counterpart; panel (b) reports SASRec-based component ablations. H and N denote HR and NDCG, respectively.}
  \label{tab:steam-results}
  \footnotesize
  \setlength{\tabcolsep}{2.2pt}
  \renewcommand{\arraystretch}{1.06}
  \begin{tabular}{@{}lcccc@{\hspace{10pt}}lrrrr@{}}
    \toprule
    \multicolumn{5}{c}{\textbf{(a) Backbone comparison}}
      & \multicolumn{5}{c}{\textbf{(b) Component ablation}} \\
    \cmidrule(lr){1-5}\cmidrule(lr){6-10}
    Backbone & H@5 & H@10 & N@5 & N@10
      & Variant & H@5 & H@10 & N@5 & N@10 \\
    \midrule
    SASRec
      & 72.0/\textbf{85.9}
      & 121.8/\textbf{141.4}
      & 45.3/\textbf{56.5}
      & 61.3/\textbf{74.3}
      & w/o Local\ Adapt.
      & 82.2 & 137.6 & 53.4 & 71.2 \\
    STEAM
      & 85.1/\textbf{96.2}
      & 148.5/\textbf{154.9}
      & 52.7/\textbf{60.3}
      & 73.0/\textbf{79.2}
      & w/o Matched Null
      & 84.0 & 139.5
      & 55.0 & 72.9 \\
    BirDRec
      & 78.3/\textbf{86.0}
      & 129.5/\textbf{134.8}
      & 50.0/\textbf{55.9}
      & 66.5/\textbf{71.6}
      & w/o Matched Ref.
      & 84.8 & 139.4 & 55.5 & 73.0 \\
    SSDRec
      & 75.5/\textbf{87.7}
      & 127.5/\textbf{145.1}
      & 50.8/\textbf{59.2}
      & 67.4/\textbf{77.4}
      & \textsc{AdaptedKG}
      & \textbf{85.9} & \textbf{141.4}
      & \textbf{56.5} & \textbf{74.3} \\
    \bottomrule
  \end{tabular}
\end{table}

\paragraph{Recommendation performance.}
\textsc{AdaptedKG} achieves higher values for every reported metric across all four backbones (Table~\ref{tab:steam-results}(a)). The gains with SASRec show that \textsc{AdaptedKG} can benefit a standard sequential recommender without an existing behavior-denoising mechanism. The gains with STEAM, BirDRec, and SSDRec suggest that its KG-derived retention signal is compatible with
behavior-denoising mechanisms learned from interaction logs.

\paragraph{Ablation study.}
In the ablations, \emph{w/o Local Adapt.} removes the entire local
adaptation stage, using the global path set with context-independent
weights. \emph{w/o Matched
Null} retains the context-specific local view but removes matched-null
calibration by setting $\alpha_p(C)=\kappa_p(C)$. \emph{w/o Matched
Ref.} replaces matched references with uniform references. All three variants lower every reported metric relative to the
full model (Table~\ref{tab:steam-results}(b)), with local adaptation
producing the largest drops and either matching stage producing smaller
declines. This pattern is consistent with local adaptation forming
example-specific path evidence, matched-null calibration refining path
selection and weighting, and matched-reference calibration refining
interaction support.

\section{Conclusion}
We presented \textsc{AdaptedKG}, which uses knowledge graph relations to
assess the contextual support for an interaction. The method compares
the observed context with structurally similar alternatives to identify
and weight paths that stand out from the background. It then uses
similar reference items to calibrate interaction support. The resulting retention coefficients adjust historical
representations and target losses during training. All graph processing is completed offline, leaving the
recommendation backbone and inference procedure unchanged. The results
show that calibrated graph evidence can complement behavior denoising
in sequential recommendation.

\section*{Declaration of Use of Generative AI}

During the preparation of this work, the authors used OpenAI Codex on a limited basis for language polishing and to assist with the implementation and debugging of experimental code. All tool-assisted text and code were critically reviewed, tested, and revised by the authors. The research design, methodological decisions, analysis, and interpretation remained the responsibility of the authors, who take full responsibility for the content of this work.

\bibliography{references}

\end{document}